\documentstyle[aps,amssymb,cite,epsfig,12pt]{revtex}
\def\bull{\vrule height .9ex width .8ex depth -.1ex } 
\def\m@th{\mathsurround=0pt }

\def\vereq#1#2{\lower3pt\vbox{\baselineskip1.5pt \lineskip1.5pt
\ialign{$\m@th#1\hfill##\hfil$\crcr#2\crcr\sim\crcr}}}
\def\gtrsim{\mathrel{\mathpalette\vereq>}}

\title{Stress Relaxation of Entangled Polymer Networks}
\author{Gary S. Grest$^a$, Mathias P\"utz$^b$, 
Ralf Everaers$^c$, and Kurt Kremer$^c$}

\address{$^a$Sandia National Laboratory, Albuquerque,
New Mexico 87185}     

\address{$^b$Center for Microengineered Materials, University
of New Mexico, Albuquerque, New Mexico 87106}

\address{$^c$Max Planck Institut f\"ur Polymerforschung,
Postfach 3148, 55021 Mainz, Germany}
\begin{document}
\maketitle
\begin{abstract}
The non-linear stress-strain relation for crosslinked polymer
networks is studied using molecular dynamics simulations.
Previously we demonstrated the importance of trapped
entanglements in determining the elastic and relaxational
properties of networks. Here we present new results for
the stress versus strain for both dry and swollen networks.
Models which limit the fluctuations of the network strands
like the tube model are shown to describe the stress for both elongation
and compression. For swollen networks, the total modulus is found to
decrease like $(V_0/V)^{2/3}$ and goes to the phantom model result
only for short strand networks.
\end{abstract}

\section{Introduction}

Understanding the elastic and relaxation properties of
crosslinked networks and rubber  has been a longstanding
problem \cite{treloar75,graessley82}. 
One of the main unresolved issues has been the 
importance of trapped entanglements in determining the
properties of the network. Edwards \cite{edwards67} first introduced
the idea that obstacles produced by other chains give rise
to a tube in which the monomers of a chain move. Later
de Gennes generalized  these same ideas to
long chains, resulting in the reptation model for the
dynamics of an entangled polymer melt \cite{degennes79,doi86}. 
However how entanglements affect the properties of a
crosslinked network remains controversial. The reason
is twofold. The first is that most experiments do not allow
good control of all the microscopic parameters. Second,
theoretical descriptions are very complicated due to the
presence of quenched disorder and usually contain several
adjustable parameters, which are difficult to relate to 
microscopic details and are almost impossible to determine
uniquely from experiment. 

Computer simulations can directly address many fundamental
questions regarding the dynamics of polymer networks. Details
of the microscopic topology, such as the elastically active
fraction of the network and loop entanglements, can be identified
and controlled. In particular, it is possible to isolate and
quantify their effects on macroscopic observables such as the
elastic modulus. Simulation results  on randomly
crosslinked, end-linked \cite{duering94c} and diamond networks
\cite{everaers95,everaers96,everaers99}  have
conclusively demonstrated the importance of trapped entanglements in 
determining the elastic and relaxational properties of the
network. Results for the mean square displacement of the
crosslinks show that they do not move completely freely
in space as assumed in the phantom network model but instead
are confined to region of space of the order of the tube diameter
of an uncrosslinked polymer melt \cite{duering94c}. Further
the monomers of strands of length $N_s>N_e$, where $N_e$ is the
entanglement length of the melt, are also confined to a tube.
The fluctuations of the middle monomers are found to diverge very
slowly \cite{duering94c}, as $N_s^{1/2}$ in accordance with reptation
theory.

To test the theoretical models, it is advantageous to study model
networks, preferably with equal strand lengths between crosslinks
and no dangling ends. As such, we have carried out elongation simulations on
ideal diamond lattice networks \cite{everaers95,everaers96,everaers99}
and end-linked networks \cite{duering94c}. In the former,
the network is constructed to have the coordination
of a diamond lattice, with a well controlled topology. A second
class of networks, which is not quite as ideal but experimentally
realizable is end-linked networks \cite{queslel84,patel92}. 
In this case one starts with a melt of linear chains and adds
$f$-functional crosslinkers to a fraction of the chain ends.
Then by changing the temperature or other conditions, these groups
can be activated, so that when
another end comes in contact a chemical reaction can occur.
We have made a number of model networks following this procedure
\cite{duering94c}. A third class which is of high practical importance
are randomly crosslinked networks, where crosslinks between melt chains
are introduced between any pair of monomers. We have also made model
systems of this type, with the difference that we attached the
end-monomers to a randomly chosen nearby monomer resulting in a
tri-functional networks instead of the usual four-functional ones.
The benefit of this method is that the networks contain no dangling
ends which slow down relaxation and are not believed to contribute
to zero-frequency stress.
Here we present new results for the non-linear stress strain
properties of end-linked networks as well as some 
results on swelling of randomly crosslinked networks.
\section{Simulation Model}
To simulate the networks we use the molecular dynamics method
which has been successfully applied to study entanglement
effects in polymer melts \cite{kremer90,puetz99a}. In this model the 
polymers are represented as freely jointed bead-spring chains. All
monomeric units of mass $m$ interact via a purely repulsive Lennard-Jones
potential
\begin{equation}
U_{IJ}(r)=
\cases{
        4\epsilon \left[ \left(\frac{\sigma}{r}\right)^{12}
                -\left(\frac{\sigma}{r}\right)^6 + \frac{1}{4} \right]
        & $r \le r_c$;\cr
        0
        & $r > r_c$, \cr
}
\label{LJ}
\end{equation}
where $r_c=2^{1/6}\sigma$. Monomer units connected along
the chain or through the crosslinking procedure are connected
by a finite extensible non-linear elastic (FENE) potential. The
model parameter are the same as in ref. \cite{kremer90}. The
temperature $T=\epsilon/k_B$. We use dimensionless units
in which $\sigma=1$ and $\epsilon=1$ and the basic unit of
time $\tau=\sigma(m/\epsilon)^{1/2}$. The simulations are
carried out at a constant density $\rho=0.85\sigma^{-3}$. 
One of the most important features of this model is that
the energy barrier for the crossing of two chain segments
is high enough ($\approx 70k_BT$) such that crossing is virtually
impossible.

In our previous studies of the mean squared displacement of the
chains in a dense melt, we found that the entanglement length
$N_e$ for this model is approximately 35 beads \cite{kremer90}.
This result is obtained by comparing the crossover time $\tau_e$
at which the motion of the monomers in the chain slows down from its
initial Rouse-like motion at early times to the slower $t^{1/4}$
behavior predicted by reptation theory. 
This result was confirmed by come recent simulations \cite{puetz99a}
on very long chains ($N=350$,$700$ and $10000$) which also found that
the segment motion of the chains was in very good agreement
with the reptation model. Estimates from these simulations
suggest $N_e=32\pm 2$ \cite{puetz99a}.  
However calculations of the plateau modulus $G_N^o$, which were not
carried out in our earlier studies due to lack of computer time,
indicate that $N_e$ is  much larger.  This estimate
of $N_{e,p}$ was obtained using the standard formula of Doi \cite{doi86}
\begin{equation}
G_N^o=\frac{4}{5} \frac{\rho k_BT}{N_{e,p}}.
\label{doi}
\end{equation}
Depending on the formula we use to fit the non-linear stress
strain data, we obtain values for $N_{e,p}$ between $65$ to $80$,
about twice as large as that obtained from the displacement of
the chains. We use the average value $72$ when comparing our
results to experiment.
This suggests an error in the pre-factor ${4\over 5}$ in
Eq.~\ref{doi}.  Experimentally $N_e$ is only determined from $G_N^o$.
We also find that diffusion
constant data from the simulations and from experiment
agree very well when both are scaled by their Rouse value for short chains 
and plotted versus $N/N_{e,p}$ \cite{puetz99a}.

Here we present results for end-linked systems $(f=4)$ consisting
of $M$ chains of length $N$, where $M/N=1000/35$, $500/100$, and $120/350$.
Due to the slow relaxation and long simulation runs required to
obtain the stress at a given strain, the results presented here are for
one configuration for $N=35$ and $100$ and two configurations for
$N=350$. For comparison we also present results for the plateau
modulus for $N=350$ and $N=700$ \cite{puetz99a}.
For details of the end crosslinking procedure, see ref. \cite{duering94c}.
For the swelling studies, we present results for randomly crosslinked
systems ($f=3$) of size $M/N/N_s=500/700/233$ and $1600/50/18$, where
$N$ is the melt precursor chain length before crosslinking and $N_s = N/3$
is the average length of a network strands between crosslinks.
We will restrict our analysis to the effect of (partial) swelling in
good solvent on the stress-strain relationship of these networks.
\section{Non-linear stress strain in the dry state}
The elastic modulus and non-linear stress versus strain 
are easily measured experimentally for networks.  Numerically
the shear modulus can be obtained from uniaxial, volume
conserving elongation of the sample which can be described by
a diagonal deformation tensor $\Lambda$ with the stretching
factor $\Lambda_{xx}=\lambda$ and the contraction factors 
$\lambda_{yy}=\Lambda_{zz}=\lambda^{-1/2}$ in the other two
directions. The normal stress
$\sigma_N$ is then readily determined from the microscopic
virial tensor, $\sigma_N=\sigma_{xx}-(\sigma_{yy}+\sigma_{zz})/2$,
where $x$ is the direction of elongation. In the present simulations
we vary $\lambda$ from $2.5$ (extension) to $0.5$ (compression). 
Due to finite system size, we are limited to relatively large
strains $\lambda\ge1.2$ or $\lambda\le 0.8$.  For smaller strains,
the stress is small and difficult to determine from the noise.
After an initial step strain, the strain decays very slowly. For the
longest chain lengths studied ($N_s=350)$, runs typically of at least
$10^5\tau$ ($\le 2\times 10^4 \tau$ for shorter network strands) were needed
to reach equilibrium.

\begin{figure}[!ht]
  \begin{center}
    \begin{tabular}{ccc}
      \resizebox*{0.46\textwidth}{!}{\includegraphics{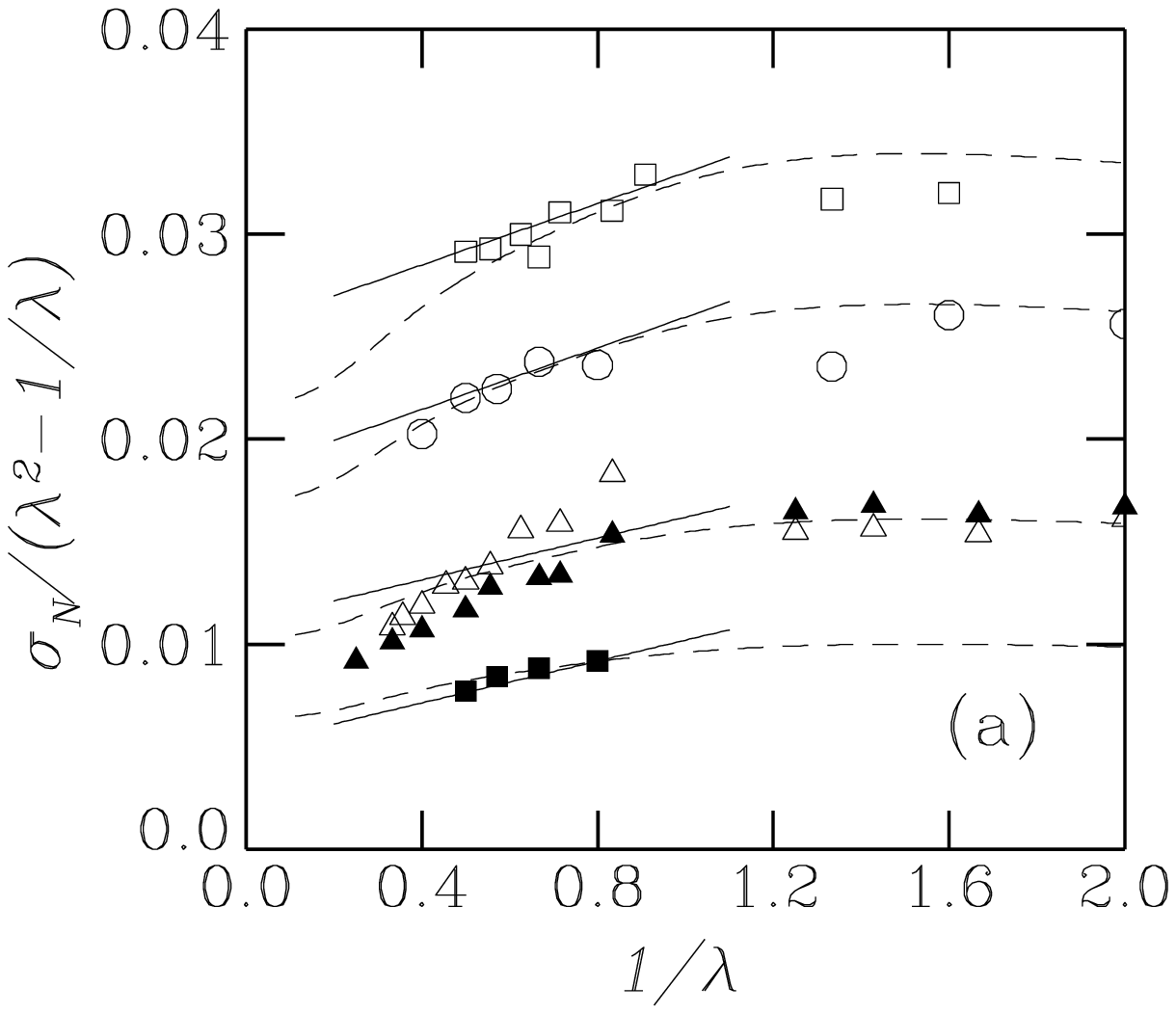}} &
      \phantom{XX} &
      \resizebox*{0.46\textwidth}{!}{\includegraphics{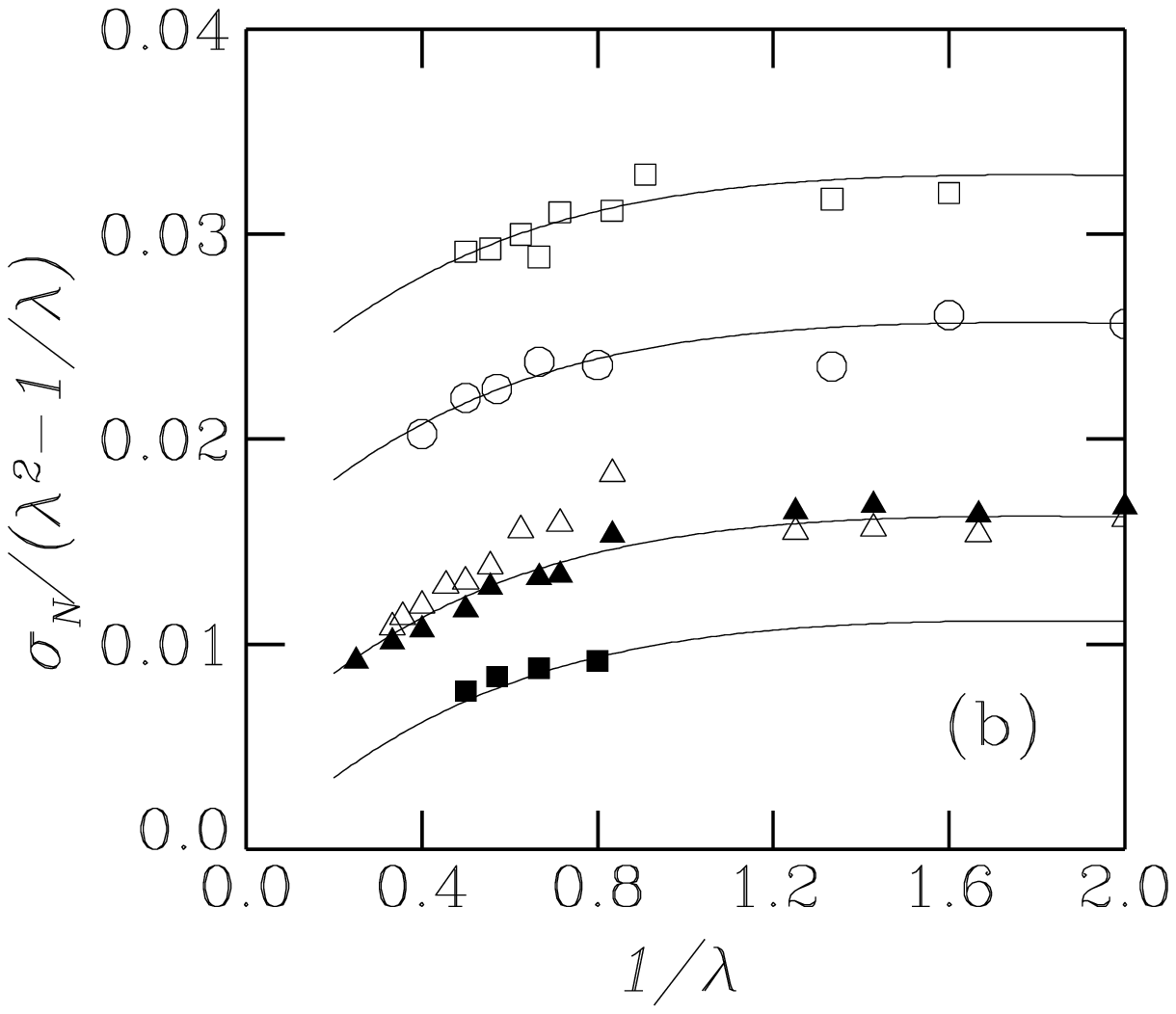}} \\
    \end{tabular}
    \caption{ \label{fig:stressstrain}
      Stress-strain relation for end crosslinked networks
      compared to (a) the Mooney-Rivlin, Eq. \ref{mrstress} (solid lines),
      and constrained junction (non-affine deformation parameter $\kappa=5.0$)
      \cite{flory77} (dashed lines) models and (b) the Rubinstein-Panyukov
      form, Eq.~(\ref{rpstress}). Data are for 
      $N_s=35\ (\Box)$, $100\ (\circ)$ and $350$ (triangles). Also
      shown is the plateau value for an uncrosslinked melt of
      chains of length $N=700\ (\bull)$.
      }
  \end{center}
\end{figure}

The classical theory of rubber elasticity \cite{treloar75} is based on
the empirical fact that the network strands have Gaussian coil statistics.
With the exterior deformation described by the tensor $\Lambda$
the resulting free energy stored in the network strands (of entirely
entropic origin) is:
\begin{equation} \label{clmodel}
  F = V \frac{k_BT}{2} \rho_{chain} \left(1-\frac{2}{f}\right)
      \sum_{i=1}^{3} \Lambda_{ii}^2,
\end{equation}
where $V$ is the volume of the sample, $\rho_{chain}$ the number density of
elastically active strands, $f$ their functionality and $\Lambda_{ii}$ are
the diagonal elements of the deformation tensor.
The above result was derived by James and Guth \cite{james4753} under
the assumptions that the crosslinks and the
connecting strands are allowed to fluctuate freely without mutual hindrance
of each other (this is commonly referred to as
the phantom chain model). If one fixes the crosslinks in space so that
they deform affinely under strain (the affine model) the factor $(1-2/f)$
disappears and on arrives at the old result given by Kuhn \cite{kuhn46}.
With this on can readily derive an expression for the normal tension 
in the sample as function of the elongation \cite{non-crystal1}, 
\begin{equation} \label{clstress}
  \sigma_N = - \frac{\lambda}{L^2} \frac{d F(\lambda)}{d\lambda}       
           = \rho_{chain} k_B T \left( 1 - \frac{2}{f} \right)
           \left( \lambda^2  - \frac{1}{\lambda} \right),
\end{equation}
where $L^2 / \lambda$ is the transversal cross section of the sample.
Flory and Erman \cite{flory77,erman78} extended the theory to allow
entanglement effects on the crosslinks by reducing their fluctuation
radii, thus interpolating between the affine and the phantom model
with a more complicated stress strain behavior. The model is quite
successful describing many experimental stress-strain relations as well
as our data (see Fig.~\ref{fig:stressstrain}a). However, due to our additional knowledge
on the conformations of the chain \cite{duering94c}, we know that the
fluctuations of the entire chains are restricted and therefore we think that
constraint models for junctions alone should be discarded.
Many models exist which limit the fluctuations of the network strands
like the tube-model by Heinrich {\it et al.} \cite{heinrich88} or
some recent constraint models by Kloczkowski {\it et al.} \cite{kloczkowski95},
Rubinstein and Panyukov (RP) \cite{rubinstein97,rubinstein99}
and Everaers \cite{everaers98}. Of these, especially the RP model
is rather appealing, since it introduces only one additional parameter,
the chain length between entanglements $N_e$. Erman's and
Everaers' recent models introduce a large set of constraint parameters
and little is known yet how to choose them, although the physical
meaning of these parameters is relatively clear.
At least for Everaers' model it has been demonstrated how to obtain
more detailed
knowledge about network fluctuations by measuring these parameters
directly through simulations of highly idealized networks with
diamond lattice topology \cite{everaers99}.

An analysis of the type presented in \cite{everaers99} is beyond
the scope of this paper and we shall limit our analysis along the
lines of the RP model which has an extra entanglement term in
the free energy:
\begin{equation} 
  F_e = \frac{1}{2} V G_e 
        \sum_{i=1}^{3} \left( \Lambda_i + \frac{1}{\Lambda_i} \right).
\end{equation}
It is customary to renormalize the stress $\sigma_N$ by the classical
functional dependence on $\lambda$,
$\sigma_N^* = \sigma_N/(\lambda^2-1/\lambda)$ to emphasize deviations
from this behavior.
Together with some recent improvement \cite{rubinstein99} on the theory
which takes into account reorganization of the constraining tubes due to
anisotropic tube deformations \cite{rubinstein97} this additional
term results in a stress-strain relation for uniaxial strain:
\begin{equation}
  \sigma_N^* = G_{c} + 1.84G_e/(\lambda+0.84\lambda^{-1/2})
\label{rpstress}
\end{equation}
In Fig.~\ref{fig:stressstrain} we show the stress-strain curves
for our various model end-linked
networks together with fits to Eq.~\ref{rpstress}
to the semi-empirical Mooney-Rivlin \cite{treloar75}
formula
\begin{equation} \label{mrstress}
  \sigma_N^* = 2C_1+2C_2/\lambda 
\end{equation}
and to the resulting stress-strain relation of the theory of Flory
and Erman (formula not shown due to complexity) for comparison.
We can see that the elongation branch can be described by each of
these equations, while the Mooney-Rivlin form provides a clearly inadequate
description of the compression data. Note, that the stress has
been normalized by the classical stress dependence (Eq.~\ref{clstress})
on $\lambda$, so that deviations from the classical behavior are
more clearly visible.

However a more quantitative analysis of the resulting fit-parameters
shows that the situation is less conclusive.
We performed two fits to the PR model. For the first fit the value of $G_e$
was determined from the plateau  modulus for the uncrosslinked melt of
free chains and was held fixed.
Using the RP form, $G_e=0.0102\epsilon/\sigma^{-3}$, while
$G_c=0.0145,\ 0.0097,$ and $0.0034\epsilon/\sigma^{-3}$ for $N_s=35$,
$100$ and $350$, respectively. If $G_c$ is equavalent to either phantom
or affine results the values of $G_c(N_s)$ should scale with the
crosslink density $(N_s^{-1})$. The values for the  ratios 
$G_c(35)/G_c(350)=4.3$ and $G_c(100)/G_c(350)=2.9$, compared to the
phantom/affine prediction of $10$ and $3.5$. The absolute values for
$G_c$ generally appear too high compared to the phantom and/or the
affine results.
A different way to look at the results is to set $G_c$ to their expected
values and fit $G_e$. For our shortest strand network $N_s=35$,
fluctuations of the crosslinks are of the order the radius
of gyration of the entire chain, 
hence we expect $G_c$ should equal its value for the phantom model,
$G_c = 0.012\epsilon/\sigma^{-3}$. Using this value for $G_c$,
we find $G_e=0.0014\epsilon/\sigma^{-3}$.
For our $N_s=350$ networks the crosslinks explore much less space
than the radius of gyration of the whole chain, thus the behavior is
expected to be more in accordance with the affine model,
$G_c=0.0022 \epsilon/\sigma^{-3}$ (after adjusting
for the fact that 10 of 120 chains are 
not elastically active). Using this value for $G_c$ we find
$G_e=0.0012\epsilon/\sigma^{-3}$. Thus the effective $G_e$ becomes
larger as the chains become shorter, i.e. the presence of more crosslinks
provides stronger constraints as in the melt case.
A more rigorous test of the RP model would be to compare networks of
strand lengths $N_s=350$ and $700$, which is beyond the amount of
available computer time at present.
However experimentally the more relevant region in parameter space 
lies in the range $N=35$ to $N=350$ since it is difficult to make
networks with very large strand lengths.

\begin{figure}[!ht]
  \begin{center}
    \resizebox*{0.7\textwidth}{!}{\includegraphics{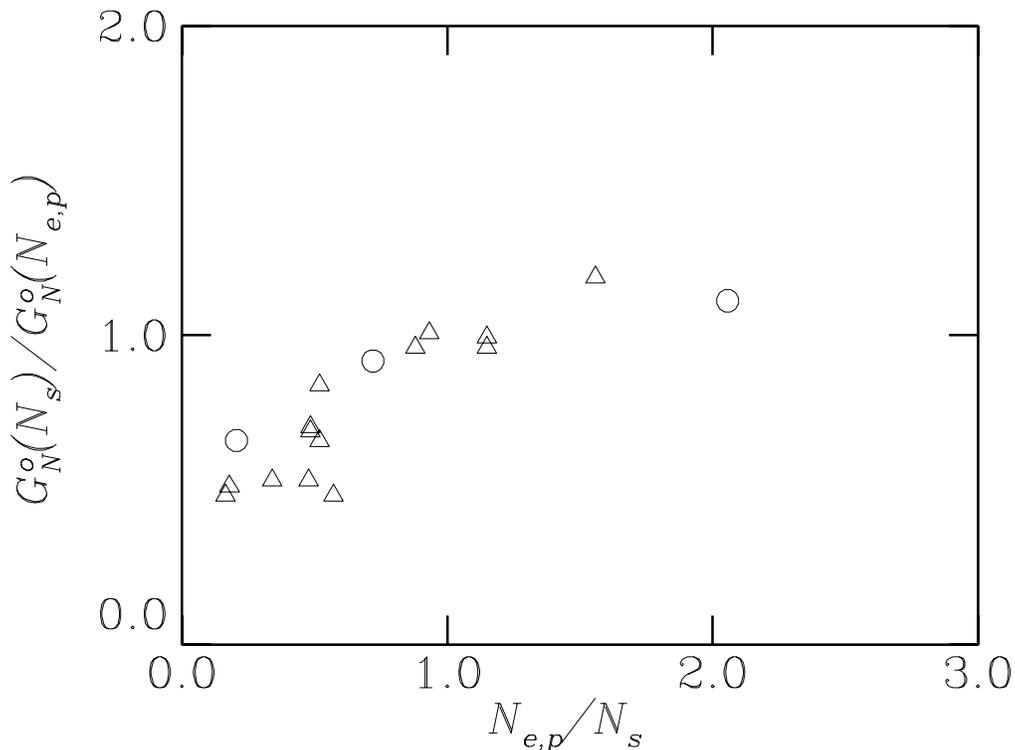}}
    \caption{ \label{fig:expstress}
      Plot of modulus for end linked networks from simulation $(\circ)$
      and experiment $(\bigtriangleup)$. The modulus has been normalized
      by its value at $N_e$.
      The experimental data for $G^o_N(M_s)$ of PDMS networks are from 
      Patel {\it et al.} \cite{patel92}. 
      }
  \end{center}
\end{figure}
The elastic modulus $G_N^o$ can be obtained from extrapolating
the slope of the non-linear stress-strain curves to $\lambda=1$.
Results for $G_N^o$ are shown in Fig.~\ref{fig:stressstrain}
using the RP form to extrapolate to $\lambda=1$. The MR form gives
larger values for $G_N^o$. To compare to experimental
data for PDMS end-linked networks, we have normalized
the modulus for its value at $N_{e,p}=72$. For PDMS the
entanglement molecular weight $M_{e,p}=9600$. As seen from
Fig.~\ref{fig:stressstrain}, our simulation results agree with
experiment. The large scatter in the experimental data for strand
lengths $N_s \gtrsim 2 \cdot N_{e,p}$ shows the difficulty to
create good networks with the endlinking method.
\section{Swollen Networks}
\begin{figure}[!htb]
  \begin{center}
    \resizebox*{0.7\textwidth}{!}{\includegraphics{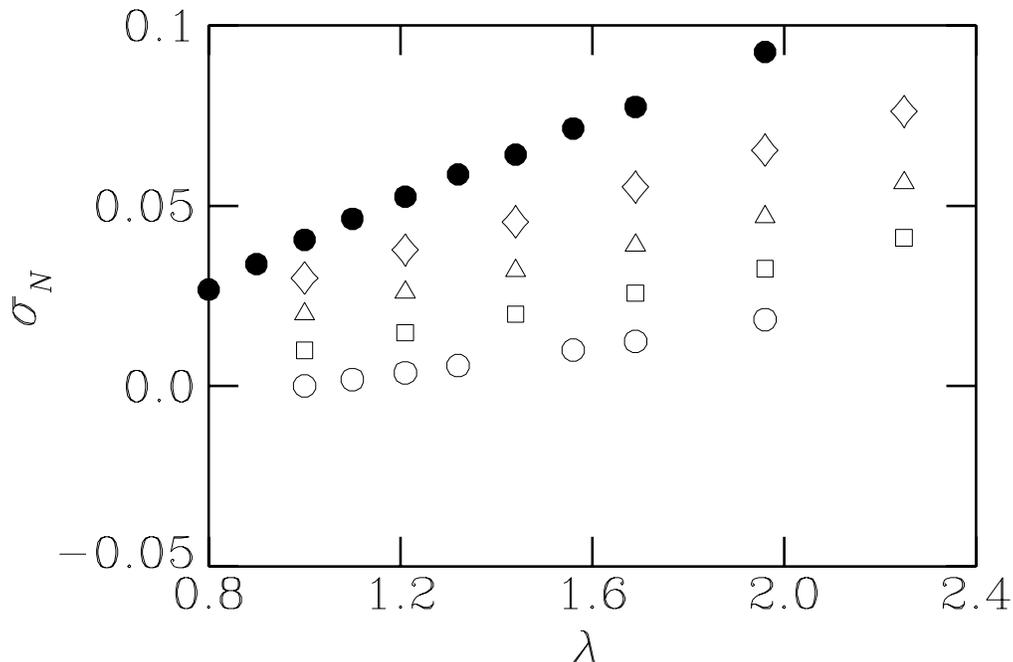}}
    \caption{ \label{fig:swellstress}
      Stress-strain relations for a randomly crosslinked
      three-functional system of average strand length $N_s=233$
      for various degrees of swelling:
      $q=1.0\ (\bullet)$, $q=1.2\ (\lozenge)$, $q=1.4\ (\vartriangle)$,
      $1=1.6\ (\Box)$, $q=1.9\ (\circ)$. For clarity the curves are
      shifted in y-direction by $0.01\epsilon/\sigma^3$ with respect
      to each other.
      }
  \end{center}
\end{figure}

Experimental results \cite{mark75,rubinstein94} show that the entanglement
contribution to the modulus decreases significantly if the networks are
swollen in good solvent. To investigate this effect we let our randomly
crosslinked networks swell in vacuum. Since our model only involves
repulsive interactions vacuum acts as a perfect solvent for our model networks.
At various degrees of (incomplete) swelling we performed stress-strain
experiments (only elongation) with two of our swollen samples.
We used formula (\ref{rpstress}) to fit the resulting set of
stress-strain curves and estimate the modulus, though  
for extracting the modulus any of the forms for $\sigma_N$ 
could have been used. Fig.~\ref{fig:swellstress} shows the stress-strain relations
for the unswollen states and at
equilibrium swelling (when elastic free energy and entropy of dilution
cancel each other). The swollen systems show the empirically
expected behavior of classical rubber elasticity (\ref{clstress})
which becomes clearer when one looks at the classical and
entanglement contributions to the modulus shown in Tab.~\ref{tab:swellmodulus}.
These were determined by fits to Eq.~(\ref{rpstress}).
One can see from the table as well as from Fig~\ref{fig:swellstress}. that
the total modulus (slope of the curves for $\lambda \to 1$) approximately
decreases like $q^{-2}=(V_0/V)^{2/3}$.
For the short strand network $N_s=18$ it is nice to see that the system
recovers the phantom result at maximum swelling (the effective values
for $G_c$ and $G_e$ are no longer meaningful, since finite chain
length effects become very important at high swelling and large
deformations which are not accounted for in the theory, but still
they are useful to extract a total modulus). For the long
chain a entanglement contribution remains, however it is shifted over
into $G_c$ which is still about a factor of $1.5$ to $4$ larger than the
affine and  phantom values, respectively).
If RP theory is applied to swelling, in the spirit of Flory and co-workers
(i.e. assuming that the statistics of the chains remains Gaussian
with swelling), this theory (as well as others) actually predicts,
that the entanglement contribution should vanish and the classical
crosslink contribution should remain constant which qualitatively
agrees with our findings.

This follows from a simple analysis of the leading powers
in $\lambda$ of the different contributions to the free energy.
Let us split up the deformation into an isotropic swelling factor
$q$ and an anisotropic deformation $\Lambda_{ii}'$ :
$\Lambda_{ii} = q \Lambda_{ii}'$.
The free energy due to crosslinks is then proportional to $q^2$, but the
system size $L$ also changes proportional to $q$. Therefore the powers
of $q$ in the formula for the stress cancel exactly. The entanglement
contribution in the RP model is proportional to the first power of $q$
only and thus its stress contribution should decrease like $1/q$,
whereas we find it behaves more like $q^{-2}$ for our networks.
The most important key to understand this behavior certainly lies in
the non-Gaussian statistics of the swollen chains. We recently found,
that the fractal structure of the chains changes significantly under swelling
and can be described as random walks only on length scales beyond the
entanglement length $N_e$ and below can be described by a fractal exponent
of $\nu \approx 0.7$ \cite{puetz99b}, which is in disagreement with the
widely assumed single chain good solvent exponent $\nu\ \simeq 3/5$.
Further our results for long-chain networks seem to favor a form
of the free energy where the entanglement contribution behaves approximately
like $F_e(q,\lambda) \to G(N_s) \cdot F_{c}(\lambda)$ for large $q$.
Loosely speaking, the trapped entanglements
behave more like physcial crosslinks for high swelling.
\begin{table}[!tb]
  \begin{center}
    \begin{tabular}{|c|c|c|c|c|c|}
      \hline
      $N_s$ & $G_{ph}$ & $q$ & $G_{c}$ & $G_{e}$ & $G_{tot}$ \\
      \hline
          &        & 1.0  & \phantom{-}0.0102(05)  & \phantom{-}0.0110(10) & \phantom{-}0.0212(15) \\
          &        & 1.2  & \phantom{-}0.0074(06)  & \phantom{-}0.0058(10) & \phantom{-}0.0132(16) \\
      233 & 0.0012 & 1.4  & \phantom{-}0.0062(05)  & \phantom{-}0.0037(02) & \phantom{-}0.0194(12) \\
          &        & 1.6  & \phantom{-}0.0059(03)  & \phantom{-}0.0018(04) & \phantom{-}0.0077(07) \\
          &        & 1.9  & \phantom{-}0.0058(05)  &           -0.0005(05) & \phantom{-}0.0053(10) \\
      \hline
          &        & 1.0  & \phantom{-}0.0190(20)  & \phantom{-}0.0120(20) & \phantom{-}0.0310(40) \\
          &        & 1.2  & \phantom{-}0.0254(60)  &           -0.0047(60) & \phantom{-}0.021(12)  \\
      18  & 0.017  & 1.3  & \phantom{-}0.0327(30)  &           -0.0166(30) & \phantom{-}0.0161(60) \\
          &        & 1.4  & \phantom{-}0.0314(12)  &           -0.0162(15) & \phantom{-}0.0151(27) \\
      \hline
    \end{tabular}
  \end{center}
  \caption{ \label{tab:swellmodulus}
    Classical crosslink ($G_{c}$) and entanglement ($G_e$) contributions
    to the total modulus $G_{tot}$ in swollen ($q^3 = V/V_0$) randomly
    crosslinked three-functional systems of average strand length $N_s$.
    Here $G_{ph}=\rho_{chain}(1-2/f)$ with $f=3$ is the modulus predicted
    by the phantom model (\ref{clmodel}). Errors in
    parenthesis are those obtained by the non-linear fit. 
    }
\end{table}
\section{Conclusion}
In this paper we have presented molecular dynamics
simulation results for the non-linear
stress strain properties of a crosslinked polymer network.
Because the topology of the network can be readily
controlled in the simulations, computer simulations
provide a powerful way to test various theoretical models.
Combined with our earlier studies, we are now beginning to obtain
a better microscopic understanding of how the topology of the
network controls the dynamics and relaxation of the chain
segments and the crosslinks. All of our data strongly
support constraint models in which the monomers are
constrained to move in a tube formed by the neighboring chains.
The crosslinking
process traps in the dynamic entanglements present in the
uncrosslinked melt. These trapped entanglements dominate
the stress relaxation and modulus when the strand length
between crosslinks is large. In the unswollen, dry state, the chains
are Gaussian in agreement with all models. However as the
system is swollen in a good solvent, they become strongly
non-Gaussian on length scales less than the distance
between crosslinks. In the dry state, a number of model
for the stress-strain relation can be used to describe the
data, although further work is necessary to arrive at a more
quantitative understanding physical of the resulting fitting parameters.
Clearly a more microscopic approach investigating network fluctuations
is necessary to distinguish between these models.
In the swollen state we find that short stranded networks eventually
recover the phantom result for the modulus, however finite
chain length corrections and a proper account for the non-Gaussian
statistics are clearly needed to descibe their entire ($\lambda \neq 1$)
stress-strain relations. 
In long strand networks the entanglement contribution to the modulus
decreases with swelling but compared to the predictions of the phantom
and affine models a significant contribution remains.
The stress strain curves at high swelling
are better described by classical (crosslinks only) theories, hence
the remaining effect of entanglements appears more like an additional
crosslink contribution.
\subsection*{Acknowledgments}
Sandia is a multiprogram laboratory operated by Sandia Corporation, a Lockheed
Martin Company, for the United States Department of Energy under Contract
DE-AC04-94AL85000.                                                             
\bibliographystyle{prsty}

\begin{thebibliography}{10}
\bibitem{non-crystal1}
Note, that most experimentalists define the normal stress as the force per
  unstretched crosssection, thus the extra factor of $\lambda$ disappears. In
  simulations we use the given definition since that corresponds to the
  definition from the microscopic virial tensor.
\end{thebibliography}

\end{document}